\documentclass[3p,times]{elsarticle}

\usepackage{ecrc}

\volume{00}

\firstpage{1}

\journalname{High Energy Density Physics}

\runauth{V. Dwarkadas et al.}

\jid{hedp}

\jnltitlelogo{High Energy Density Physics}

\CopyrightLine{2011}{Published by Elsevier Ltd.}

\usepackage{amssymb}

\usepackage[figuresright]{rotating}

\begin{document}
\newcommand{\vper}{\mbox{${v_{\perp}}$}}
\newcommand{\vpar}{\mbox{${v_{\parallel}}$}}
\newcommand{\uper}{\mbox{${u_{\perp}}$}}
\newcommand{\vperout}{\mbox{${{v_{\perp}}_{o}}$}}
\newcommand{\uperout}{\mbox{${{u_{\perp}}_{o}}$}}
\newcommand{\vperin}{\mbox{${{v_{\perp}}_{i}}$}}
\newcommand{\uperin}{\mbox{${{u_{\perp}}_{i}}$}}
\newcommand{\upar}{\mbox{${u_{\parallel}}$}}
\newcommand{\uparout}{\mbox{${{u_{\parallel}}_{o}}$}}
\newcommand{\vparout}{\mbox{${{v_{\parallel}}_{o}}$}}
\newcommand{\uparin}{\mbox{${{u_{\parallel}}_{i}}$}}
\newcommand{\vparin}{\mbox{${{v_{\parallel}}_{i}}$}}
\newcommand{\dout}{\mbox{${\rho}_{o}$}}
\newcommand{\din}{\mbox{${\rho}_{i}$}}
\newcommand{\da}{\mbox{${\rho}_{1}$}}
\newcommand{\mfast}{\mbox{$\dot{M}_{f}$}}
\newcommand{\mslow}{\mbox{$\dot{M}_{a}$}}
\newcommand{\beqn}{\begin{eqnarray}}
\newcommand{\eeqn}{\end{eqnarray}}
\newcommand{\be}{\begin{equation}}
\newcommand{\ee}{\end{equation}}
\newcommand{\noi}{\noindent}
\newcommand{\ftheta}{\mbox{$f(\theta)$}}
\newcommand{\gtheta}{\mbox{$g(\theta)$}}
\newcommand{\ltheta}{\mbox{$L(\theta)$}}
\newcommand{\stheta}{\mbox{$S(\theta)$}}
\newcommand{\utheta}{\mbox{$U(\theta)$}}
\newcommand{\xitheta}{\mbox{$\xi(\theta)$}}
\newcommand{\vs}{\mbox{${v_{s}}$}}
\newcommand{\ro}{\mbox{${R_{0}}$}}
\newcommand{\pa}{\mbox{${P_{1}}$}}
\newcommand{\va}{\mbox{${v_{a}}$}}
\newcommand{\vo}{\mbox{${v_{o}}$}}
\newcommand{\vp}{\mbox{${v_{p}}$}}
\newcommand{\vw}{\mbox{${v_{w}}$}}
\newcommand{\vf}{\mbox{${v_{f}}$}}
\newcommand{\lprime}{\mbox{${L^{\prime}}$}}
\newcommand{\uprime}{\mbox{${U^{\prime}}$}}
\newcommand{\sprime}{\mbox{${S^{\prime}}$}}
\newcommand{\xiprime}{\mbox{${{\xi}^{\prime}}$}}
\newcommand{\mdot}{\mbox{$\dot{M}$}}
\newcommand{\msun}{\mbox{$M_{\odot}$}}
\newcommand{\yr}{\mbox{${\rm yr}^{-1}$}}
\newcommand{\kms}{\mbox{${\rm km} \;{\rm s}^{-1}$}}
\newcommand{\lambdav}{\mbox{${\lambda}_{v}$}}
\newcommand{\lequ}{\mbox{${L_{eq}}$}}
\newcommand{\eqpratio}{\mbox{${R_{eq}/R_{p}}$}}
\newcommand{\ra}{\mbox{${r_{o}}$}}
\newcommand{\bfig}{\begin{figure}[h]}
\newcommand{\efig}{\end{figure}}
\newcommand{\tone}{\mbox{${t_{1}}$}}
\newcommand{\done}{\mbox{${{\rho}_{1}}$}}
\newcommand{\dsn}{\mbox{${\rho}_{SN}$}}
\newcommand{\dzero}{\mbox{${\rho}_{0}$}}
\newcommand{\ve}{\mbox{${v}_{e}$}}
\newcommand{\vej}{\mbox{${v}_{ej}$}}
\newcommand{\Mch}{\mbox{${M}_{ch}$}}
\newcommand{\mej}{\mbox{${M}_{e}$}}
\newcommand{\Mst}{\mbox{${M}_{ST}$}}
\newcommand{\dam}{\mbox{${\rho}_{am}$}}
\newcommand{\Rst}{\mbox{${R}_{ST}$}}
\newcommand{\Vst}{\mbox{${V}_{ST}$}}
\newcommand{\Tst}{\mbox{${T}_{ST}$}}
\newcommand{\no}{\mbox{${n}_{0}$}}
\newcommand{\Efif}{\mbox{${E}_{51}$}}
\newcommand{\rsh}{\mbox{${R}_{sh}$}}
\newcommand{\msh}{\mbox{${M}_{sh}$}}
\newcommand{\vsh}{\mbox{${V}_{sh}$}}
\newcommand{\vrev}{\mbox{${v}_{rev}$}}
\newcommand{\rpr}{\mbox{${R}^{\prime}$}}
\newcommand{\mpr}{\mbox{${M}^{\prime}$}}
\newcommand{\vpr}{\mbox{${V}^{\prime}$}}
\newcommand{\tpr}{\mbox{${t}^{\prime}$}}
\newcommand{\cone}{\mbox{${c}_{1}$}}
\newcommand{\ctwo}{\mbox{${c}_{2}$}}
\newcommand{\cthree}{\mbox{${c}_{3}$}}
\newcommand{\cfour}{\mbox{${c}_{4}$}}
\newcommand{\Te}{\mbox{${T}_{e}$}}
\newcommand{\Ti}{\mbox{${T}_{i}$}}
\newcommand{\Ha}{\mbox{${H}_{\alpha}$}}
\newcommand{\Rprime}{\mbox{${R}^{\prime}$}}
\newcommand{\Vprime}{\mbox{${V}^{\prime}$}}
\newcommand{\Tprime}{\mbox{${T}^{\prime}$}}
\newcommand{\Mprime}{\mbox{${M}^{\prime}$}}
\newcommand{\rprime}{\mbox{${r}^{\prime}$}}
\newcommand{\rfprime}{\mbox{${r}_f^{\prime}$}}
\newcommand{\vprime}{\mbox{${v}^{\prime}$}}
\newcommand{\tprime}{\mbox{${t}^{\prime}$}}
\newcommand{\mprime}{\mbox{${m}^{\prime}$}}
\newcommand{\Me}{\mbox{${M}_{e}$}}
\newcommand{\nh}{\mbox{${n}_{H}$}}
\newcommand{\rr}{\mbox{${R}_{2}$}}
\newcommand{\rf}{\mbox{${R}_{1}$}}
\newcommand{\vtwo}{\mbox{${V}_{2}$}}
\newcommand{\vout}{\mbox{${V}_{1}$}}
\newcommand{\dshell}{\mbox{${{\rho}_{sh}}$}}
\newcommand{\dwind}{\mbox{${{\rho}_{w}}$}}
\newcommand{\dslow}{\mbox{${{\rho}_{s}}$}}
\newcommand{\dfast}{\mbox{${{\rho}_{f}}$}}
\newcommand{\vfast}{\mbox{${v}_{f}$}}
\newcommand{\vslow}{\mbox{${v}_{s}$}}
\newcommand{\cc}{\mbox{${\rm cm}^{-3}$}}
\newcommand{\apj}{\mbox{ApJ}}
\newcommand{\apjl}{\mbox{ApJL}}
\newcommand{\apjs}{\mbox{ApJS}}
\newcommand{\aj}{\mbox{AJ}}
\newcommand{\araa}{\mbox{ARAA}}
\newcommand{\nat}{\mbox{Nature}}
\newcommand{\aap}{\mbox{AA}}
\newcommand{\gca}{\mbox{GeCoA}}
\newcommand{\pasp}{\mbox{PASP}}
\newcommand{\mnras}{\mbox{MNRAS}}
\newcommand{\apss}{\mbox{ApSS}}
\newcommand{\memsai}{\mbox{MMSaI}}

\begin{frontmatter}



\dochead{}

\title{Supernova Remnant Evolution in Wind Bubbles: A Closer Look at
  Kes 27}


\author[vvd]{V. V. Dwarkadas\corref{cor1}}
\ead{vikram@oddjob.uchicago.edu}

\author[dd]{D. Dewey}
\ead{dd@space.mit.edu}

\address[vvd]{Department of Astronomy and Astrophysics, University of
  Chicago, TAAC 55, Chicago, IL 60637}

\address[dd]{MIT Kavli Institute, Cambridge MA 02139 }

\begin{abstract}

Massive Stars ($> 8 \msun$ ) lose mass in the form of strong winds.
These winds accumulate around the star, forming wind-blown bubbles.
When the star explodes as a supernova (SN), the resulting shock wave
expands within this wind-blown bubble, rather than the interstellar
medium.  The properties of the resulting remnant, its dynamics and
kinematics, the morphology, and the resulting evolution, are shaped by
the structure and properties of the wind-blown bubble. In this article
we focus on Kes 27, a supernova remnant (SNR) that has been proposed
by \cite{cssl08} to be evolving in a wind-blown bubble, explore its
properties, and investigate whether the properties could be ascribed
to evolution of a SNR in a wind-blown bubble. Our initial model does
not support this conclusion, due to the fact that the reflected shock
is expanding into much lower densities.

\end{abstract}


\end{frontmatter}


\section{Introduction}
\label{sec:intro}

About 80\% of supernovae (SNe) arise from the core collapse of massive
stars. These stars experience strong mass-loss in the form of winds,
modifying the medium around them and creating low density wind blown
bubbles surrounded by dense shells \cite[see also Dwarkadas \&
  Rosenberg HEDLA 2012 proceedings]{Weaver1977, ta11}. When the
supernova remnant (SNR) shock interacts with the dense shell, it
results in a transmitted shock that expands slowly into the denser
medium, and a reflected shock that initially travels back into the
already-shocked lower-density medium.  The reflected shock continues
to travel inward and eventually passes the reverse shock, speeding its
progress into the SN ejecta.

The kinematic and morphological properties of the SNR would be
determined by its expansion in the bubble
\cite{dwarkadas11}. Expansion of a SNR in a wind bubble has been
studied by many authors \cite{cl89, tbfr90, trfb91, dwarkadas2005,
  dwarkadas2007a, dwarkadas2007b}. One of the most famous examples of
a SNR expanding in a pre-existing wind-blown cavity is the nearby SN
1987A, which has been extensively studied \cite{cd95, mccray07,
  dwarkadas07c, deweyetal12}. Although many, if not most,
core-collapse SNe should be evolving in wind-blown cavities, it is
difficult to find many clear observational examples, and therefore to
compare the theoretical models to observations. In this paper we wish
to study the SNR Kesteven 27, which has been suggested as being an
example of a SNR in a wind-blown cavity.

\section{Kes 27}
\label{sec:kes27}
\begin{figure}[ht]
\includegraphics[scale=0.8]{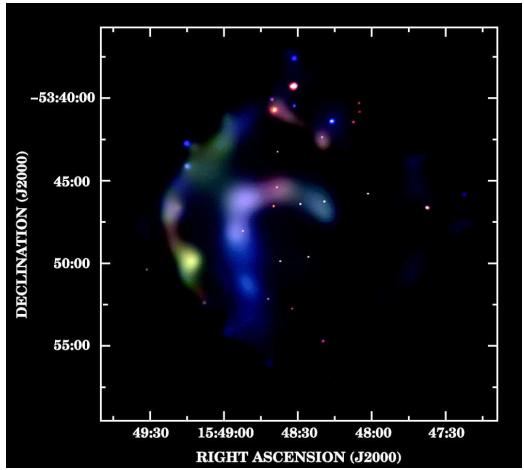}
\caption{Tricolor ACIS-I image of Kes 27. Taken from \cite{cssl08},
  reproduced by permission of the AAS. The X-ray intensities in the
  0.3-1.5, 1.5-2.2, and 2.2-7.0 keV bands are coded red, green, and
  blue, respectively, and are scaled logarithmically in the ranges
  19-170, 19-290, and 27-270 photons cm$^{-2} s^{¡V1}
  sr^{¡V1}$.  \label{fig:kes27arcs}}
\end{figure}

Kesteven 27 is a member of the class of thermal composite or
mixed-morphology remnants \cite{rp98}, which show thermal X-ray
emission extending a long way in to the center. X-ray data has been
obtained using {\it Einstein}, {\it ROSAT}, {\it ASCA} and {\it
  Chandra}. In the radio band it has been observed at 843 MHz with
{\it MOST} \cite{wg96}. The radio emission is brightest towards the
East, but fades towards the Northwest.  The Chandra image (Fig
\ref{fig:kes27arcs}, \cite{cssl08}) show two incomplete shell-like
features in the northeastern half, with brightness fading towards the
southwest. The X-ray and radio structure led Chen et al.~2008
\cite[hereafter CSSL08]{cssl08} to suggest that the morphology
represents a SNR expanding in a wind-blown bubble, as shown in Fig
\ref{fig:kes27arcs}.  In their view, the remnant is colliding with an
HI shell to the northeast, sending a reflected shock back into the
interior.  The two circular arcs were assumed to be representative of
emission from near the location of transmitted and reflected shock
waves. Calculations based on equations by \cite{sgro75} were used to
quantify their claims.

\section{Numerical Simulations}
\label{sec:maxen}

We have carried out numerical hydrodynamic simulations to understand
the structure of Kesteven 27, with the intention of exploring the
reflected shock model suggested by \cite{cssl08}.  The simulations
were carried out using the VH-1 finite difference hydrodynamic code in
2D spherical co-ordinates.  Radiative cooling was included via a
cooling function. Further details of the numerical method are
described in earlier papers \cite{db98, dwarkadas2005, dwarkadas2007b}

\begin{figure}[ht]
\includegraphics[scale=0.7]{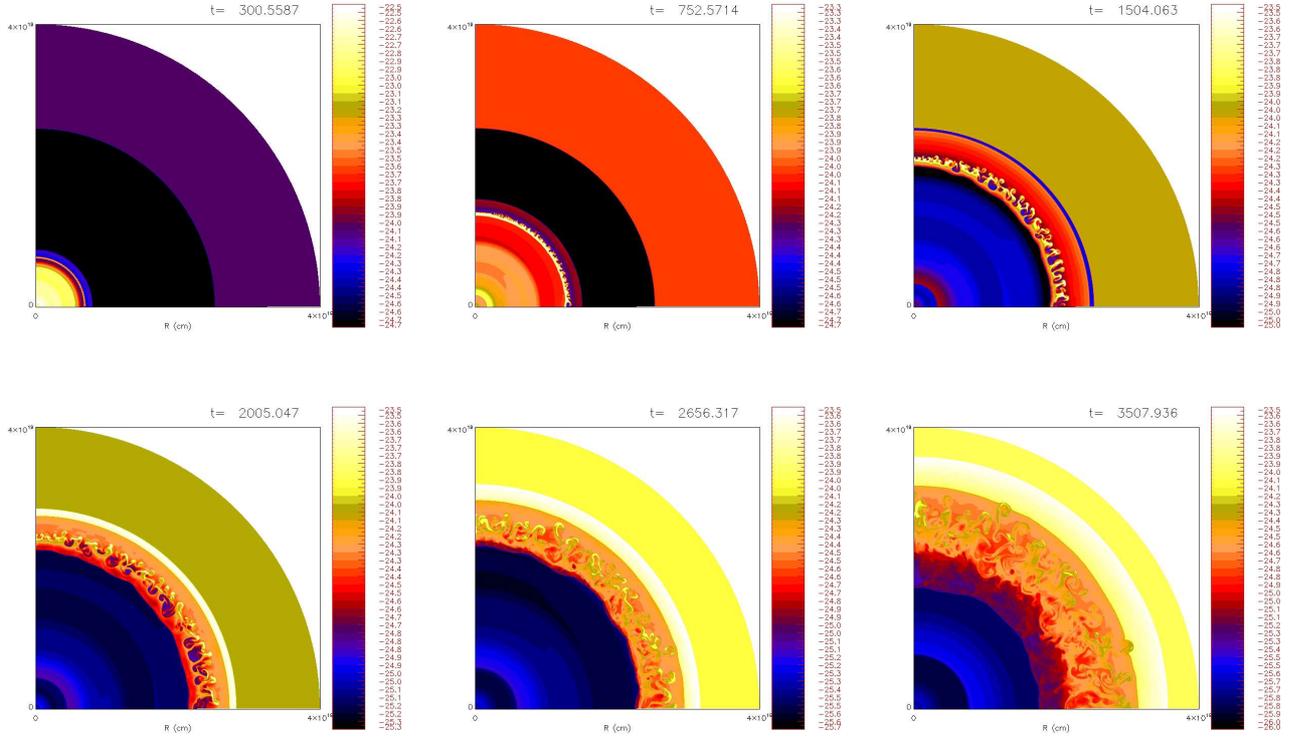}
\caption{(1) Evolution of the SNR shock waves in a low density medium
  (2) Growth of Rayleigh-Taylor (R-T) instabilities is visible at the
  decelerating contact discontinuity.  (3) The instabilities have
  grown in size as the forward shock approaches the HI shell.  (4)
  Collision with the denser HI shell drives a slow-moving shock into
  the shell, and gives rise to a reflected shock moving back into the
  ejecta.  (5) Transmitted and reflected shocks advance, while the R-T
  unstable region continues to grow. As the forward shock has slowed
  down considerably, the instabilities reach almost to the forward
  shock, which would not normally be the case. (6) Reflected and
  transmitted shocks have radii consistent with observations.
  However, as we show in the paper, emission from the reflected shock
  is very low compared to that from the forward shocked
  region. \label{fig:kes27new}}
\end{figure}

Our basic model has the following parameters: A power-law ejecta
density structure as defined in \cite{chevalier1982a}, with ejecta
density assumed to decrease as r$^{-9}$ \cite{chevalier1994}, and an
ejecta mass of 4.5 $\msun$. The shock first expands into a low density
medium (n$_H$=0.1), stretching for about 8.4 pc, then collides with a
higher density HI shell, with n$_H$ =0.5. In Fig~\ref{fig:kes27new},
we show density snapshots from the evolution.  Time in years is given
at the top of each plot. Initially (Fig~\ref{fig:kes27new}, top row),
the expansion of the SN ejecta into the medium gives rise to a forward
shock expanding into the medium and a reverse shock going back into
the ejecta, separated by a contact discontinuity. As the shock expands
into the low density medium, the decelerating contact discontinuity is
seen to become Rayleigh-Taylor unstable, as expected
\cite{dwarkadas2000}. After about 1500 years, the forward shock
collides with the denser HI shell, giving rise to a slowly expanding
shock transmitted into the HI shell, and a faster moving reflected
shock that moves much faster through the low density interior
(Fig~\ref{fig:kes27new}, bottom row).  At about 3500 years the radii
of the transmitted and reflected shocks are approximately consistent
with the positions of the observed arcs.  Thus our models are able to
reproduce the gross morphology suggested by CSSL08.

\begin{figure}[ht]
\includegraphics[scale=0.75]{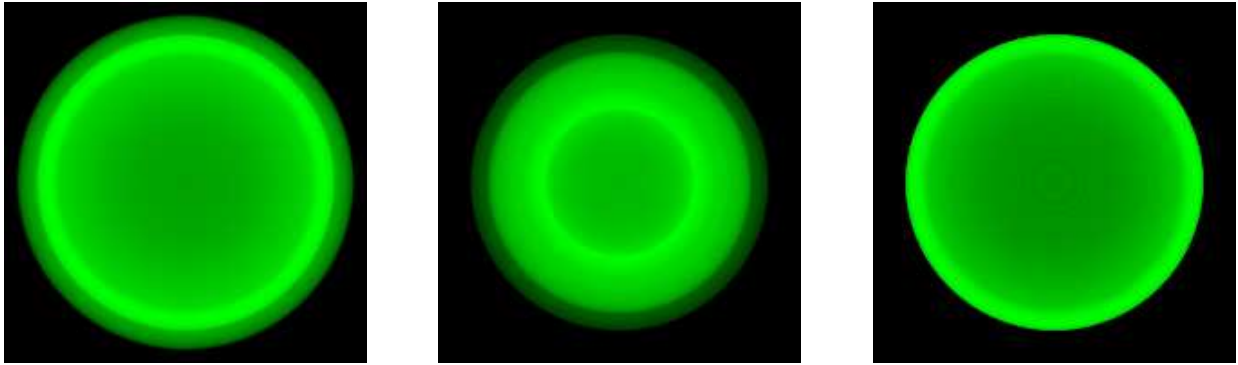}
\caption{ Simulated images based on the hydrodynamics.  The radial
  average properties were used to make 3D images (using ISIS/V3D)
  assuming full spherical emission; the intensity scale is non-linear
  to better show the full range.  The images correspond to the density
  (left), the X-ray emissivity for constant density (middle), and,
  finally, the X-ray emission taking into account the density-squared
  dependence (right).  As can be seen, in projection only the outer
  ring is manifest, and hardly any emission is seen in the inner
  regions. The visualization does not produce two visible arcs as seen
  in Fig \ref{fig:kes27arcs}. \label{fig:simim} }
\end{figure}

To explore whether the X-ray morphology and emission is consistent
with the observations, we have made preliminary calculations of these
quantities from the simulations. Averaging over radial profiles, we
have extracted mean radial profiles of density and temperature from
the 2D simulations.  The temperature is used to compute the radial
X-ray emissivity (flux per density-squared) by evaluating a simple
non-equilibrium X-ray emission model ({\tt vnei}) with
$T_e=0.2\,T_{\rm hydro}$, $n_et=3\times 10^{11}$~s\,cm$^{-3}$, and
abundances determined from the X-ray data.  Using the ISIS \cite{hd00}
and V3D \cite{dn09} packages, we convert the radial values into
simulated 3D images for several quantities from the last timestep. We
find that, unlike the observed data, the simulated X-ray image
(Fig~\ref{fig:simim}, right image) in this case shows only one outer
ring of emission and does not show any inner arc coincident with the
reflected shock. Although hot X-ray emitting gas is certainly present
at the inner locations (Fig~\ref{fig:kes27new} middle image), the
reason this inner ring is not visible is that the inner shock is
traveling into a much lower density region (Fig~\ref{fig:kes27new}
left image). Since X-ray emissivity is proportional to the square of
the density, the $\geq$ 5 times lower density results in a decrease in
X-ray flux by a factor of over 25, and thus no appreciable amount of
emission is produced in the interior.

\section{Discussion}

As shown in the above simulations, we conclude that using the
parameters generally defined in CSSL08, it is difficult to reproduce
the observed X-ray morphology of Kes 27.

Is it likely that a different set of parameters would work? It seems
unlikely in general as long as the reflected shock is traveling into
the low density ejecta. In our simulations the reflected shock is
already expanding into the constant density plateau of the
ejecta. While it may be supposed that the emissivity could be higher
if it were still expanding in the steep power-law part of the ejecta
density, it is unlikely that this would work in practice for a remnant
that is a few thousand years old. This is because the ejecta density
decreases with time as t$^{-3}$, and therefore no matter which part of
the density profile it was interacting with, it would have decreased
to a low value in about 3500 years. If the collision happened much
earlier in the SNR history, and the ejecta mass was significantly
larger, then it is possible that the parameters could be fine-tuned so
that the reflected shock would be expanding in a high-enough density
regime to provide appreciable emission.

But overall this appears unlikely. The model requires that the SNR
collide with a high density shell in order to produce a reflected
shock in the first place. This means that the density of the material
into which the forward shock expands must by necessity be larger than
that of the plasma in which the reflected shock is expanding. It is
then difficult to come up with a set of conditions where the reflected
shock finds itself in a region where the density is high enough to
provide emissivity comparable to that of the forward shock. It is
possible that a high metallicity in the low density region could raise
the emissivity, and a really high density in the ambient medium could
slow the forward shock down to the extent that its temperature, and
therefore X-ray emissivity, is quite low. However these are extremes
that do not easily fit into the scenario suggested by CSSL08, and it
would be extremely difficult to produce a viable model with these
characteristics.

It is hard to think of circumstances where the reverse shock is
expanding into a denser medium than the forward shock is. One case
that we have recently examined \cite{ddb10}, for the very young SN
1996cr, may be able to provide some inspiration. This SN appears to
have interacted with a dense shell of gas very early (a couple of
years) after explosion. After about 7 years, the SN shock has crossed
the dense shell and is expanding in a lower density medium, whereas
the reverse shock is still in the higher density shell. The emission
predominantly then arises from a reverse shock. If such a situation
could be envisaged for an older SNR, where the forward shock has
crossed a dense shell and is expanding in lower density material,
whereas the reverse shock is still expanding in the higher density
material, then it is possible that we could reproduce two arcs of
emission.

In future we will carry out a more thorough parameter survey to study
the remnant expansion, keeping the above factors in mind. We also plan
improvements to our techniques to allow us to use the full
multi-dimensional simulations in calculations of the X-ray emission,
and production of simulated images.

\subsection{Relation to High Energy Physics Lab Experiments} In keeping 
with the theme of lab astrophysics at this conference, we suggest that
this is an astrophysical scenario that is quite feasible to be carried
out with available apparatus \cite{drakeetal00, kleinetal03}, and thus
to be studied under controlled settings in the laboratory. Basically
one needs to create an environment which is partially filled with
medium 1 of a given density, and partly with medium 2 that has density
5 times higher. A strong shock expands first in medium 1, and then
collides with medium 2, which has the higher density. The collision
would lead to the formation of a transmitted and reflected shock at
the interface. The structure and expansion of these shocks must be
studied over a few doubling times. All shocks would be
non-radiative. This would allow a very interesting and potentially
significant astrophysical scenario to be studied in the laboratory,
which could answer pertinent questions about shock propagation in wind
bubbles and other astrophysical environments.

\section{Acknowledgments} 
VVD's research is funded by grant TM1-12005X, provided by NASA through
the Chandra X-ray Observatory center (CXC), operated by SAO under NASA
contract NAS8-03060. DD's research is supported by SAO contract
SV3-73016 to MIT for support of the CXC and Science Instruments.












\end{document}